\documentclass[journal]{IEEEtran}
\usepackage{lineno}
\usepackage{amsmath}
\usepackage{multirow}
\usepackage{subcaption}
\usepackage{relsize}
\usepackage{slashbox}
\usepackage{cite}

\ifCLASSINFOpdf
\usepackage[pdftex]{graphicx}
\DeclareGraphicsExtensions{.pdf,.jpeg,.png}
\else
\fi

\begin{document}

\title{Light Gated Recurrent Units  \\ for Speech Recognition}

        
\author{Mirco~Ravanelli,
        Philemon~Brakel,
        Maurizio~Omologo,
        Yoshua~Bengio
        }

\markboth{IEEE Journal of EMERGING TOPICS IN COMPUTATIONAL INTELLIGENCE}%
{Shell \MakeLowercase{\textit{et al.}}: Bare Demo of IEEEtran.cls for IEEE Journals}
\maketitle
\footnote{© 2018 IEEE. Personal use of this material is permitted. Permission from IEEE must be obtained for all other uses, in any current or future media, including reprinting/republishing this material for advertising or promotional purposes, creating new collective works, for resale or redistribution to servers or lists, or reuse of any copyrighted component of this work in other works.  DOI: 10.1109/TETCI.2017.2762739 URL: \url{http://ieeexplore.ieee.org/document/8323308/}}
\begin{abstract}
A field that has directly benefited from the recent advances in deep learning is Automatic Speech Recognition (ASR). 
Despite the great achievements of the past decades, however, a natural and robust human-machine speech interaction still appears to be out of reach, especially in challenging environments characterized by significant noise and reverberation.
To improve robustness, modern speech recognizers often employ acoustic models based on Recurrent Neural Networks (RNNs), that are naturally able to exploit large time contexts and long-term speech modulations. It is thus of great interest to continue the study of proper techniques for improving the effectiveness of RNNs in processing speech signals.

In this paper,  we revise one of the most popular RNN models, namely Gated Recurrent Units (GRUs), and propose a simplified architecture that turned out to be very effective for ASR. The contribution of this work is two-fold: First, we analyze the role played by the reset gate, showing that a significant redundancy with the update gate occurs. As a result, we propose to remove the former from  the GRU design, leading to a more efficient and compact single-gate model. Second, we propose to replace hyperbolic tangent with ReLU activations. This variation couples well with batch normalization and could help the model learn long-term dependencies without numerical issues.

Results show that the proposed architecture, called Light GRU (Li-GRU), not only reduces the per-epoch training time by more than 30\% over a standard GRU, but also  consistently improves the recognition accuracy across different tasks, input features, noisy conditions, as well as across different ASR paradigms, ranging from standard DNN-HMM speech recognizers to end-to-end CTC models.
\end{abstract}


\IEEEpeerreviewmaketitle

\section{Introduction}
Deep learning is an emerging technology that is considered one of the most promising directions for reaching higher levels of artificial intelligence \cite{Goodfellow-et-al-2016-Book}. 
This paradigm is rapidly evolving and some noteworthy achievements of the last years include, among the others, the development of effective regularization methods \cite{dropout,batchnorm}, improved optimization algorithms \cite{adam}, and  better architectures \cite{cnn1,tdnn2,lstm_highway,gru1}. The exploration of generative models \cite{gan}, deep reinforcement learning \cite{alpha_go} as well as the evolution of sequence to sequence paradigms \cite{dima_nmt} also represent important milestones in the field. Deep learning is now being deployed in a wide range of domains, including bio-informatics, computer vision, machine translation, dialogue systems and natural language processing, just to name a few.
Another field that has been transformed by this technology is Automatic Speech Recognition (ASR) \cite{lideng}.  Thanks to modern Deep Neural Networks (DNNs), current speech recognizers are now able to significantly outperform previous GMM-HMM systems, allowing ASR to be applied in several contexts, such as web-search, intelligent personal assistants, car control and radiological reporting.

Despite the progress of the last decade, state-of-the-art speech recognizers are still far away from reaching satisfactory robustness and flexibility. This lack of robustness typically happens  when facing challenging acoustic conditions \cite{adverse}, characterized by considerable levels of non-stationary noise and acoustic reverberation \cite{dsr_summary,pawel2,hain,dnn_rev,dnn_rev2,dnn3,ravanelli_SLT,ravanelli15,ravanelli_icassp}. 
The development of robust ASR has been recently fostered by the great success of some international challenges such as CHiME \cite{chime3}, REVERB \cite{revch_short} and ASpIRE \cite{aspire}, which were also extremely useful to establish common evaluation frameworks among researchers.


Currently, the dominant approach to automatic speech recognition relies on a combination of a discriminative DNN and a generative Hidden Markov Model (HMM). The DNN is normally employed for acoustic modeling purposes to predict context-dependent phone targets. The acoustic-level predictions are later embedded in an HMM-based framework, that also integrates phone-transitions, lexicon, and language model information to retrieve the final sequence of words.
An emerging alternative is end-to-end speech recognition, that aims to drastically simplify the current ASR pipeline by using fully discriminative systems that learn everything from data without (ideally) any additional human effort. Popular end-to-end techniques are attention models and Connectionist Temporal Classification (CTC).
Attention models are based on an encoder-decoder architecture coupled with an attention mechanism \cite{attention1} that decides which input information to analyze at each decoding step.
CTC \cite{CTC_graves} is based on a DNN predicting symbols from a predefined alphabet (characters, phones, words) to which an extra unit (\textit{blank}) that emits no labels is added.
Similarly to HMMs, the likelihood (and its gradient with respect to the DNN parameters) are computed with dynamic programming by summing over all the paths that are possible realizations of the ground-truth label sequence. This way, CTC allows one to optimize the likelihood of the desired output sequence directly, without the need for an explicit label alignment.

For both the aforementioned frameworks, Recurrent Neural Networks (RNNs) represent a valid alternative to standard feed-forward DNNs. RNNs, in fact, are more and more often employed in speech recognition, due to their capabilities to properly manage time contexts and capture long-term speech modulations.

In the machine learning community, the research of novel and powerful RNN models is a very active research topic. General-purpose RNNs such as Long Short Term Memories (LSTMs) \cite{lstm} have been the subject of several studies and modifications over the past years \cite{peephole, lstm_odyssey, lstm_highway}. This evolution has recently led to a novel architecture called Gated Recurrent Unit (GRU) \cite{gru1}, that simplifies the complex LSTM cell design.

Our work continues these efforts by further revising GRUs. 
Differently from previous efforts, our primary goal is not to derive a general-purpose RNN, but to modify the standard GRU design in order to better address speech recognition.
In particular, the major contribution of this paper is twofold: First, we propose to remove the reset gate from the network design. Similarly to \cite{mgru}, we found that removing it does not significantly affect the system performance, also due to a certain redundancy observed between update and reset gates.
Second, we propose to replace hyperbolic tangent (tanh) with Rectified Linear Unit (ReLU) activations \cite{relu} in the state update equation. 
ReLU units have been shown to be more effective than sigmoid non-linearities for feed-forward DNNs \cite{krizhevsky2012imagenet,dahl2012context}.
Despite its recent success, this non-linearity has largely been avoided for RNNs, due to the numerical instabilities caused by the unboundedness of ReLU activations: composing many times GRU layers (i.e., GRU units following an affine transformation) with sufficiently large weights can lead to arbitrarily large state values. However, when coupling our ReLU-based GRU with batch normalization \cite{batchnorm}, we did not experience such numerical issues. This allows us to take advantage of both techniques, that have been proven effective to mitigate the vanishing gradient problem as well as to speed up network training.    

We evaluated our proposed architecture on different tasks, datasets, input features, noisy conditions as well as on different ASR frameworks (i.e., DNN-HMM and CTC). Results show that the revised architecture reduces the per-epoch training wall-clock time by more than 30\%, while improving the recognition accuracy.  Moreover, the proposed solution leads to a compact model, that is arguably easier to interpret, understand and implement, due to a simplified design based on a single gate. 

The rest of the paper is organized as follows. Sec.~\ref{sec:rev_gru} recalls the standard GRU architecture, while Sec.~\ref{sec:novel} illustrates in detail the proposed model and the related work.
In Sec.~\ref{sec:setup}, a description of the adopted corpora and experimental setup is provided. The results are then reported in Sec.~\ref{sec:res}. Finally, our conclusions are drawn in Sec.~\ref{sec:conc}.

\section{Gated Recurrent Units} \label{sec:rev_gru}

The most suitable architecture able to learn short and long-term speech dependencies is represented by RNNs \cite{lideng}. RNNs, indeed, can potentially capture temporal information in a very dynamic fashion, allowing the network to freely decide the amount of contextual information to use for each time step. Several works have already highlighted the effectiveness of RNNs in various speech processing tasks, such as speech recognition \cite{graves,lstm_speech,baidu,joint6,chime4_paper}, speech enhancement \cite{dnn_se3}, speech separation \cite{sep_lstm,ndnn1} as well as speech activity detection \cite{lstm_vad}.
Training RNNs, however, can be complicated by vanishing and exploding gradients, that might impair learning long-term dependencies \cite{Bengio94}.
Although exploding gradients can be tackled with simple clipping strategies \cite{pascanau}, the vanishing gradient problem requires special architectures to be properly addressed. A common approach relies on the so-called gated RNNs, whose core idea is to introduce a gating mechanism for better controlling the flow of the information through the various time-steps. Within this family of architectures, vanishing gradient issues are mitigated by creating effective ``shortcuts", in which the gradients can  bypass multiple temporal steps.

The most popular gated RNNs are LSTMs \cite{lstm}, that often achieve state-of-the-art performance in several machine learning tasks, including speech recognition \cite{graves,lstm_speech,baidu,dnn_se3,joint6,chime4_paper}.
LSTMs rely on memory cells that are controlled by forget, input, and output gates.
Despite their effectiveness, such a sophisticated gating mechanism might result in an overly complex model. On the other hand, computational efficiency is a crucial issue for RNNs and considerable research efforts have recently been devoted to the development of alternative architectures \cite{lstm_odyssey,gru3,lstm_highway}.

A noteworthy attempt to simplify LSTMs has recently led to a novel model called Gated Recurrent Unit (GRU) \cite{gru1,gru2}, that is based on just two multiplicative gates. 
In particular, the standard GRU architecture is defined by the following equations: 

\begin{subequations}
\begin{align}
z_{t}&=\sigma(W_{z}x_{t}+U_{z}h_{t-1}+b_{z}), \\
\label{eq:eq_2}r_{t}&=\sigma(W_{r}x_{t}+U_{r}h_{t-1}+b_{r}), \\
\label{eq:eq_3}\widetilde{h_{t}}&=\tanh(W_{h}x_{t}+U_{h}(h_{t-1} \odot r_{t})+b_{h}), \\
\label{eq:eq_4}h_{t}&=z_{t} \odot h_{t-1}+ (1-z_{t}) \odot \widetilde{h_{t}}.
\end{align}
\end{subequations}

where $z_{t}$ and $r_{t}$ are vectors corresponding to the update and reset gates, respectively, while $h_{t}$ represents the state vector for the current time frame $t$.
Element-wise multiplications are denoted with $\odot$.
The activations of both gates are logistic sigmoid functions $\sigma(\cdot)$, that constrain $z_{t}$ and $r_{t}$ to take values ranging from 0 and 1. The candidate state $\widetilde{h_{t}}$ is processed with a hyperbolic tangent. 
The network is fed by the current input vector $x_{t}$ (e.g., a vector of speech features), while the parameters of the model are the matrices $W_z$, $W_r$, $W_h$ (the feed-forward connections) and $U_z$, $U_r$, $U_h$ (the recurrent weights).
The architecture finally includes trainable bias vectors $b_z$, $b_r$ and $b_h$, that are added before the non-linearities are applied. 

As shown in Eq.~\ref{eq:eq_4}, the current state vector $h_{t}$ is a linear interpolation between the previous activation $h_{t-1}$ and the current candidate state $\widetilde{h_{t}}$. The weighting factors are set by the update gate $z_{t}$, that decides how much the units will update their activations. This linear interpolation is the key component for learning long-term dependencies. If $z_{t}$ is close to one, in fact, the previous state is kept unaltered and can remain unchanged for an arbitrary number of time steps. On the other hand, if $z_{t}$ is close to zero, the network tends to favor the candidate state $\widetilde{h_{t}}$, that depends more heavily on the current input and on the closer hidden states. The candidate state $\widetilde{h_{t}}$ also depends on the reset gate $r_{t}$, that allows the model to possibly delete the past memory by forgetting the previously computed states.

\section{A novel GRU framework} \label{sec:novel}
The main changes to the standard GRU model concern the reset gate, ReLU activations, and batch normalization, as outlined in the next sub-sections. 

\subsection{Removing the reset gate}
From the previous introduction to GRUs, it follows that the reset gate can be useful when significant discontinuities occur in the sequence. For language modeling, this may happen when moving from one text to another that is not semantically related. In such situation, it is convenient to reset the stored memory to avoid taking a decision biased by an unrelated history. 
 \begin{figure}[t!]
 \centering
 \includegraphics[width=0.45\textwidth]{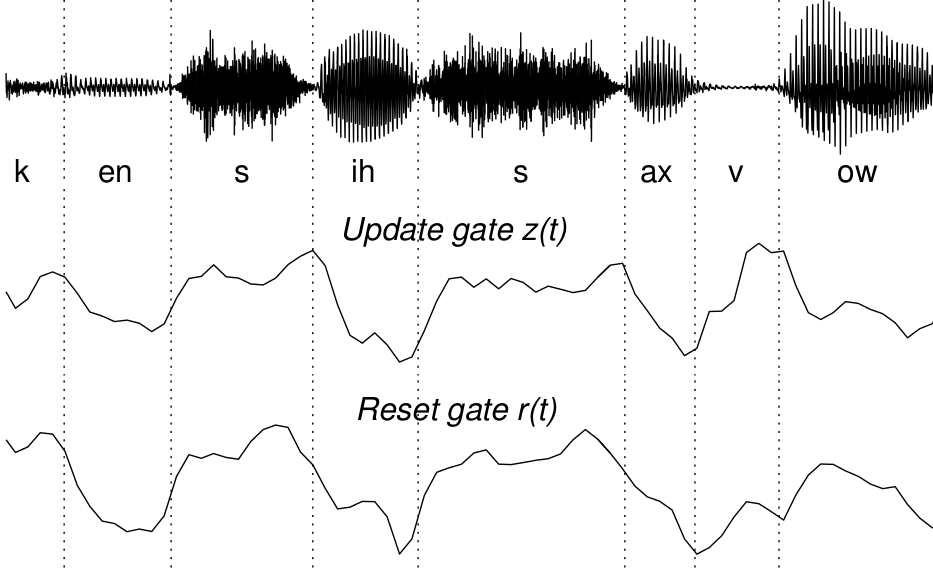}
 \caption{Average activations of the update and reset gates for a GRU trained on TIMIT in a chunk of the utterance ``\textit{sx403}" of speaker ``\textit{faks0}".}
 \label{fig:im1}
 \end{figure}

Nevertheless, we believe that for some specific tasks like speech recognition this functionality might not be useful. 
In fact, a speech signal is a sequence that evolves rather slowly (the features are typically computed every 10 ms), in which the past history can virtually always be helpful.  
Even in the presence of strong discontinuities, for instance observable at the boundary between a vowel and a fricative, completely resetting the past memory can be harmful. On the other hand, it is helpful to memorize phonotactic features, since some phone transitions are more likely than others.

We also argue that a certain redundancy in the activations of reset and update gates might occur when processing speech sequences.  For instance, when it is necessary to give more importance to the current information,  the GRU model can set small values of $r_{t}$. A similar effect can be achieved with the update gate only, if small values are assigned to $z_{t}$. The latter solution tends to weight more the candidate state $\widetilde{h_{t}}$, that depends heavily on the current input. 
Similarly, 
a high value can be assigned either to $r_{t}$ or to $z_{t}$, in order to place more importance on past states. 
This redundancy is also highlighted in Fig. \ref{fig:im1}, where a temporal correlation in the average activations of update and reset gates 
can be readily appreciated for a GRU trained on TIMIT. 
This degree of redundancy will be analyzed in a quantitative way in Sec. \ref{sec:res} using the cross-correlation metric $C(z,r)$:
\begin{equation}
C(z,r)=\overline{z}_t \star \overline{r}_t
\label{eq:cross}
\end{equation}
where $\overline{z}_t$ and $\overline{r}_t$ are the average activations (over the neurons) of update and reset gates, respectively, and $\star$ is the cross-correlation operator.

Based on these reasons, the first variation to standard GRUs thus concerns the removal of the reset gate $r_{t}$. 
This change leads to the following modification of Eq. \ref{eq:eq_3}:

\begin{equation}
\widetilde{h_{t}}=\tanh(W_{h}x_{t}+U_{h} h_{t-1}+b_{h})
\end{equation}

The main benefits of this intervention are related to the improved computational efficiency, that is achieved thanks to a more compact single-gate model.

\subsection{ReLU activations}
The second modification consists in replacing the standard hyperbolic tangent with ReLU activation. In particular, we modify the computation of candidate state $\widetilde{h_{t}}$ (Eq.~\ref{eq:eq_3}), as follows:

\begin{equation}
\widetilde{h_{t}}=\mbox{ReLU}(W_{h}x_{t}+U_{h}h_{t-1}+b_{h})
\end{equation}

Standard tanh activations are less used in feedforward networks because they do not work as well as piecewise-linear activations when training deeper networks~\cite{xavier}. The adoption of ReLU-based neurons, that have shown to be effective in improving such limitations, was not so common in the past for RNNs. This was due to numerical instabilities originating from the unbounded ReLU functions applied over long time series. However, coupling this activation function with batch normalization turned out to be helpful for taking advantage of ReLU neurons without numerical issues, as will be discussed in the next sub-section.




\subsection{Batch Normalization} \label{sec:bn}
Batch normalization \cite{batchnorm} has been recently proposed in the machine learning community and addresses the so-called \textit{internal covariate shift} problem by normalizing the mean and the variance of each layer's pre-activations for each training mini-batch. Several works have already shown that this technique is effective both to improve the system performance and to speed-up the training procedure \cite{cesar,tim,baidu,ravanelli_SLT,ravanelli_icassp}. Batch normalization can be applied to RNNs in different ways. In \cite{cesar}, the authors suggest to apply it to feed-forward connections only, while in \cite{tim} the normalization step is extended to recurrent connections, using separate statistics for each time-step.  In our work, we tried both approaches, but  we did not observe substantial benefits when extending batch normalization to recurrent parameters (i.e., $U_{h}$ and $U_{z}$).  For this reason, we applied this technique to feed-forward connections only (i.e., $W_{h}$ and $W_{z}$), obtaining a more compact model that is almost equally performing but significantly less computationally expensive. When batch normalization is limited to feed-forward connections, indeed, all the related computations become independent at each time step and they can be performed in parallel. This offers the possibility to apply it with reduced computational efforts.  As outlined in the previous sub-section, coupling the proposed model with batch-normalization \cite{batchnorm} could also help in limiting the numerical issues of ReLU RNNs. Batch normalization, in fact, rescales the neuron pre-activations, inherently bounding the values of the ReLU neurons. 
In this way, our model concurrently takes advantage of the well-known benefits of both ReLU activation and batch normalization.
In our experiments, we found that the latter technique helps against numerical issues also when it is limited to feed-forward connections only.

Formally, removing the reset gate, replacing the hyperbolic tangent function with the ReLU activation, and applying batch normalization, now leads to the following model:

\begin{subequations}
\begin{align}
\label{eq:eq_5a}&z_{t}=\sigma(BN(W_{z}x_{t})+U_{z}h_{t-1}), \\
\label{eq:eq_5b}&\widetilde{h_{t}}=\mbox{ReLU}(BN(W_{h}x_{t})+U_{h}h_{t-1}), \\
\label{eq:eq_5c}&h_{t}=z_{t} \odot h_{t-1}+ (1-z_{t}) \odot \widetilde{h_{t}}.
\end{align}
\end{subequations}
The batch normalization $BN(\cdot)$ works as described in \cite{batchnorm}, and is defined as follows:

\begin{equation}
BN(a)=\gamma \odot
\label{eq:bn} \frac{a-\mu_b}{\sqrt[]{\sigma_b^2+\epsilon}}+\beta
\end{equation}
where $\mu_b$ and $\sigma_b$ are the minibatch mean and variance, respectively. A small constant $\epsilon$ is added for numerical stability. The variables $\gamma$ and $\beta$ are trainable scaling and shifting parameters, introduced to restore the network capacity. Note that the presence of $\beta$ makes the biases $b_h$ and $b_z$ redundant. Therefore, they are omitted in Eq. \ref{eq:eq_5a} and \ref{eq:eq_5b}.

We called this architecture Light GRU (Li-GRU), to emphasize the simplification process conducted on a standard GRU.

\subsection{Related work} \label{sec:related_work}
A first attempt to remove $r_{t}$ from GRUs has recently led to a single-gate architecture called Minimal Gated Recurrent Unit (M-GRU) \cite{mgru}, that achieves a performance comparable to that obtained by standard GRUs in handwritten digit recognition as well as in a sentiment classification task.  To the best of our knowledge, our contribution is the first attempt that explores this architectural variation in speech recognition.  Recently, some attempts have also been done for embedding ReLU units in the RNN framework. For instance, in \cite{orth_init} authors replaced tanh activations with ReLU neurons in a vanilla RNN, showing the capability of this model to learn long-term dependencies when a proper orthogonal initialization is adopted. In this work, we extend the use of ReLU to a GRU architecture.

In summary, the novelty of our approach consists in the integration of three key design aspects (i.e, the removal of the reset gate, ReLU activations and batch normalization) in a single model, that turned out to be particularly suitable for speech recognition. 
The potential benefits Li-GRUs have been preliminarily observed as part of a work on speech recognition described in \cite{ravanelli_is17}. This study extends our previous effort in several ways. First of all, we better analyze the correlation arising between reset and update gates. We then analyze some gradient statistics, and we better study the impact of batch normalization. 
Moreover, we assess our approach on a larger variety of speech recognition tasks, considering several different datasets as well as noisy and reverberant conditions. Finally, we extend our experimental validation to a end-to-end CTC model.

\section{Experimental setup} \label{sec:setup}
In the following sub-sections, the considered corpora, the RNN setting as well as the HMM-DNN and CTC setups are described.

\subsection{Corpora and tasks}
\label{sec:corpora}
The main features of the corpora considered in this work are summarized in Tab. \ref{tab:corpora}.

\begin{table}[t!]
\centering
\tabcolsep=0.17cm
    \begin{tabular}{ | l | c | c | c | c | }
    \cline{1-5}
   {\backslashbox{\em{Features}}{\em{Dataset.}}} & TIMIT &  DIRHA & CHiME & TED \\ \hline
Hours & 5 & 12 & 75 & 166\\ \hline
Speakers   & 462 & 83 & 87 & 820\\ \hline
Sentences & 3696 & 7138 & 43690 & 272191\\ 
\hline
\end{tabular}
\caption{Main features of the training datasets adopted in this work.}
\label{tab:corpora}
\end{table}

A first set of experiments with the TIMIT corpus was performed to test the proposed model in a close-talking scenario. These experiments are based on the standard phoneme recognition task, which is aligned with that proposed in the Kaldi s5 recipe \cite{kaldi}.

To validate our model in a more realistic scenario, a set of experiments was also conducted in distant-talking conditions  with the DIRHA-English corpus\footnote{This dataset is being distributed by the Linguistic Data Consortium (LDC).} \cite{dirha_asru}. The reference context was a domestic environment characterized by the presence of non-stationary noise (with an average SNR of about 10dB) and acoustic reverberation (with an average reverberation time $T_{60}$ of about 0.7 seconds).
Training was based on the original WSJ-5k corpus (consisting of 7138 sentences uttered by 83 speakers) that was contaminated with a set of impulse responses measured in a real apartment.
The test phase was carried out with both real and simulated datasets, each consisting of 409 WSJ sentences uttered by six native American speakers. A development set of 310 WSJ sentences uttered by six different speakers was also used for hyperparameter tuning.
To test our approach in different reverberation conditions, other contaminated versions of the latter training and test data are generated with different impulse responses. These simulations are based on the image method \cite{image} and correspond to four different reverberation times $T_{60}$ (ranging from 250 to 1000 ms). More details on the realistic impulse responses adopted in this corpus can be found in \cite{rav_is16,ravanelli_eusipco2012,rav_in14}.

Additional experiments were conducted with the CHiME 4 dataset \cite{chime4_paper}, that is based on both real and simulated data recorded in four noisy environments (on a bus, cafe, pedestrian area, and street junction). The training set is composed of 43690 noisy WSJ sentences recored by five microphones (arranged on a tablet) and uttered by a total of 87 speakers. The development set (DT) is based on 3280 WSJ sentences uttered by four speakers (1640 are real utterances referred to as DT-real, and 1640 are simulated denoted as DT-sim). 
The test set (ET) is based on  1320 real utterances (ET-real) and 1320 simulated sentences (DT-real) from other four speakers. The experiments reported in this paper are based on the single channel setting, in which the test phase is carried out with a single microphone (randomly selected from the considered microphone setup). 
More information on CHiME data can be found in \cite{chime3}.

To evaluate the proposed model on a larger scale ASR task, some  additional experiments were performed with the TED-talk dataset, that was released in the context of the IWSLT evaluation campaigns \cite{iwslt_2011}. The training set is composed of 820 talks with a total of about 166 hours of speech. The development test is composed of 81 talks (16 hours), while the test sets (TST 2011 and TST 2012) are based on 8 talks (1.5 hours) and 32 talks (6.5 hours), respectively.

\subsection{RNN setting} \label{sec:rnn_setup}
The architecture adopted for the experiments consisted of multiple recurrent layers, that were stacked together prior to the final softmax classifier. These recurrent layers were bidirectional RNNs \cite{graves}, which were obtained by concatenating the forward hidden states (collected by processing the sequence from the beginning to the end) with backward hidden states (gathered by scanning the speech in the reverse time order).
Recurrent dropout was used as regularization technique. Since extending standard dropout to recurrent connections hinders learning long-term dependencies, we followed the approach introduced in \cite{drop_asru,Gal2016}, that tackles this issue by sharing the same dropout mask across all the time steps. Moreover,  batch normalization was adopted exploiting the method suggested in \cite{cesar}, as discussed in Sec.~\ref{sec:rev_gru}.
The feed-forward connections of the architecture were initialized according to the \textit{Glorot}'s scheme \cite{xavier}, while recurrent weights were initialized with orthogonal matrices \cite{orth_init}. Similarly to \cite{ravanelli_SLT}, the gain factor $\gamma$ of batch normalization was initialized to 0.1 and the shift parameter $\beta$ was initialized to 0.

Before training, the sentences were sorted in ascending order according to their lengths and, starting from the shortest utterance, minibatches of 8 sentences were progressively processed by the training algorithm.
This sorting approach minimizes the need of zero-paddings when forming mini-batches, resulting helpful to avoid possible biases on batch normalization statistics. 
Moreover, the sorting approach exploits a curriculum learning strategy \cite{curriculum} that has been shown to slightly improve the performance and to ensure numerical stability of gradients. The optimization was done using the Adaptive Moment Estimation (Adam) algorithm \cite{adam} running for 22 epochs (35 for the TED-talk corpus) with $\beta_1=0.9$, $\beta_2=0.999$, $\epsilon=10^{-8}$. The performance on the development set was monitored after each epoch, while the learning rate was halved when the performance improvement went below a certain threshold ($th=0.001$). Gradient truncation was not applied, allowing the system to learn arbitrarily long time dependencies.

The main hyperparameters of the model (i.e., learning rate, number of hidden layers, hidden neurons per layer, dropout factor) were optimized on the development data. 
In particular, we guessed some initial values according to our experience, and starting from them we performed a grid search to progressively explore better configurations. A total of 20-25 experiments were conducted for all the various RNN models.

\subsection{DNN-HMM setup} \label{sec:hmm_dnn}
In the DNN-HMM experiments, the DNN is trained to predict context-dependent phone targets. 
The feature extraction is based on blocking the signal into frames of 25 ms with an overlap of 10 ms.  The experimental activity is conducted considering different acoustic features, i.e., 39 MFCCs (13 static+$\Delta$+$\Delta\Delta$), 40 log-mel filter-bank features (FBANKS), as well as 40 fMLLR features (extracted as reported in the s5 recipe of Kaldi \cite{kaldi}).

The labels were derived by performing a forced alignment procedure on the original training datasets. See the standard s5 recipe of Kaldi for more details \cite{kaldi}.
During test, the posterior probabilities generated  for each frame by the RNN are normalized by their prior probabilities. 
The obtained likelihoods are processed by an HMM-based decoder, that, after integrating the acoustic, lexicon and language model information in a single search graph, finally estimates the sequence of words uttered  by the speaker.
The RNN part of the ASR system was implemented with Theano \cite{theano}, that was  coupled with the Kaldi decoder \cite{kaldi} to form a context-dependent RNN-HMM speech recognizer\footnote{The code is available at \url{http://github.com/mravanelli/theano-kaldi-rnn/}.}.

\subsection{CTC setup} \label{sec:ctc_setup}
The models used for the CTC experiments consisted of 5 layers of bidirectional RNNs of either 250 or 465 units. Unlike in the other experiments, weight noise was used for regularization. The application of weight noise is a simplification of adaptive weight noise \cite{graves2011practical} and has been successfully used before to regularize CTC-LSTM models \cite{graves2013speech}. The weight noise was applied to all the weight matrices and sampled from a zero mean normal distribution with a standard deviation of $0.01$. Batch normalization was used with the same initialization settings as in the other experiments. Glorot's scheme was used to initialize all the weights (also the recurrent ones). The input features for these experiments were 123 dimensional FBANK features (40 + energy + $\Delta$+$\Delta\Delta$). These features were also used in the original work on CTC-LSTM models for speech recognition \cite{graves2013speech}.
The CTC layer itself was trained on the 61 label set. Decoding was done using the best-path method \cite{CTC_graves}, without adding any external phone-based language model.
After decoding, the labels were mapped to the final 39 label set. The models were trained for 50 epochs using batches of 8 utterances. 


\section{Results} \label{sec:res}
In the following sub-sections, we describe the experimental activity conducted to assess the proposed model. Most of the experiments reported in the following are based on hybrid DNN-HMM speech recognizers, since the latter ASR paradigm typically reaches state-of-the-art performance. However, for the sake of comparison, we also extended the experimental validation to an end-to-end CTC model.
More precisely, in sub-section \ref{sec:corr}, we first  quantitatively analyze the correlations between the update and reset gates in a standard GRU. In sub-section \ref{sec:grad}, we extend our study with some analysis of gradient statistics. The role of batch normalization and the CTC experiments are described in sub-sections \ref{sec:bn_exp} and \ref{sec:cts}, respectively. The speech recognition performance will then be reported for TIMIT,  DIRHA-English, CHiME as well as for the TED-talk corpus in  sub-sections ~\ref{sec:timit}, \ref{sec:dirha}, \ref{sec:chime}, and \ref{sec:ted_talks}, respectively. The computational benefits of Li-GRU are finally discussed in sub-section ~\ref{sec:tr_time}. 

\subsection{Correlation analysis} \label{sec:corr}
\begin{figure}[t]
\centering
  \includegraphics[scale=0.50]{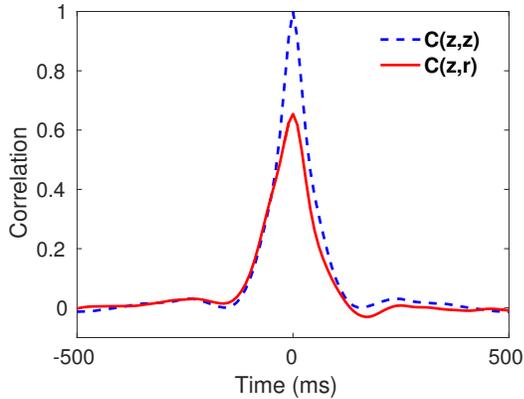}
\caption{Auto-correlation $C(z,z)$ and cross-correlation $C(z,r)$ between the average activations of the update (z) and reset (r) gates. Correlations are normalized by the maximum of $C(z,z)$ for graphical convenience.}\label{fig:corr}
\end{figure}

The correlation between update and reset gates have been preliminarily discussed in Sec. \ref{sec:rev_gru}. In this sub-section, we take a step further by analyzing it in a quantitative way using the cross-correlation function defined in Eq. \ref{eq:cross}.
In particular, the cross-correlation $C(z,r)$ between the average activations of update $z$ and reset $r$ gates is shown in Fig.~\ref{fig:corr}.
The gate activations are computed for all the input frames and, at each time step, an average over the hidden neurons is considered.
The cross-correlation $C(z,r)$ is displayed along with the auto-correlation $C(z,z)$, that represents the upper-bound limit of the former function.  

Fig. \ref{fig:corr} clearly shows a high peak of $C(z,r)$, revealing that update and reset gates end up being  redundant. This peak is about 66\% of the maximum of $C(z,z)$ and it is centered at $t=0$, indicating that almost no-delay occurs between gate activations.  This result is obtained with a single-layer GRU of 200 bidirectional neurons fed with MFCC features and trained with TIMIT. After the training-step, the cross-correlation is averaged over all the development sentences.

\begin{figure}[t]
\centering
  \includegraphics[scale=0.50]{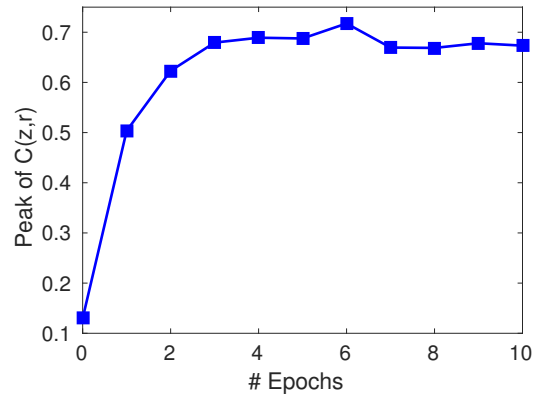}
\caption{Evolution of the  peak of the cross-correlation $C(z,r)$ over various training epochs.}\label{fig:corr2}
\end{figure}

It would be of interest to examine the evolution of this correlation over the epochs. With this regard, Fig. \ref{fig:corr2} reports the peak of $C(z,r)$ for some training epochs, showing that the GRU attributes rather quickly a similar role to update and reset gates. 
In fact, after 3-4 epochs, the correlation peak reaches its maximum value, that is almost maintained for all the subsequent training iterations.   

\subsection{Gradient analysis} \label{sec:grad}
The analysis of the main gradient statistics can give some preliminary  indications on the role played by the various parameters. 

With this goal, Table \ref{tab:grad} reports the L2 norm of the gradient for the main parameters of the considered GRU models.  
\begin{table}[t!]
\centering
\tabcolsep=0.20cm
    \begin{tabular}{ | c | c | c | c | c | }
    \cline{1-4}
   {\backslashbox{\em{Param.}}{\em{Arch.}}} & GRU &  M-GRU & Li-GRU \\ \hline
$\|W_{h}\|$ & 4.80 & 4.84 & 4.86 \\ \hline
$\|W_{z}\|$ & 0.85 & 0.89 & 0.99 \\ \hline
$\|W_{r}\|$ & 0.22 & - & - \\ \hline
$\|U_{h}\|$ & 0.15 & 0.35 & 0.71 \\ \hline
$\|U_{z}\|$ & 0.05 & 0.11 & 0.13 \\ \hline
$\|U_{r}\|$ & 0.02 & - & - \\ \hline
\end{tabular}
\caption{$L_2$ norm of the gradient for the main parameters of the GRU models. The norm is averaged over all the training sentences and epochs.}
\label{tab:grad}
\end{table}
Results reveal that the reset gate weight matrices (i.e., $W_{r}$ and $U_{r}$) have a smaller gradient norm when compared to the other parameters. 
This result is somewhat expected, since the reset gate parameters are processed by two different non-linearities (i.e., the sigmoid of Eq. \ref{eq:eq_2} and the tanh of \ref{eq:eq_3}), that can attenuate their gradients.
This would anyway indicate that, on average, the reset gate has less impact on the final cost function, further supporting its removal from the GRU design. When avoiding it, the norm of the gradient tends to increase (see for instance the recurrent weights $U_{h}$ of M-GRU model). 
This suggests that the functionalities of the reset gate, can be performed by other model parameters.    
The norm further increases in the case of Li-GRUs, due to the adoption of ReLu units. This non-linearity, indeed, improves the back-propagation of the gradient over both time-steps and hidden layers, making long-term dependencies easier to learn.
Results are obtained with the same GRU used in subsection \ref{sec:corr}, considering M-GRU and Li-GRU models with the same number of hidden units.

\subsection{Role of batch normalization} \label{sec:bn_exp}
After the preliminary analysis on correlation and gradients done in previous sub-sections, we now compare RNN models in terms of their final speech recognition performance. 
To highlight the importance of batch normalization,  Table \ref{tab:bn} compares the Phone Error Rate (PER\%) achieved with and without this technique.

\begin{table}[t!]
\centering
\tabcolsep=0.20cm
    \begin{tabular}{ | l | c | c | c | c | }
    \cline{1-4}
   {\backslashbox{\em{Param.}}{\em{Arch.}}} & GRU &  M-GRU & Li-GRU \\ \hline
without batch norm & 18.4 & 18.6 & 20.4 \\ \hline
with batch norm & 17.1 & 17.2 & 16.7 \\ \hline
\end{tabular}
\caption{PER(\%) of the GRU models with and without batch normalization (TIMIT dataset, MFCC features).}
\label{tab:bn}
\end{table}

Results show that batch normalization is helpful to improve the ASR performance, leading to a relative improvement of about 7\% for GRU and M-GRU and 18\% for the proposed Li-GRU.
The latter improvement confirms that our model couples particularly well with this technique, due to the adopted ReLU activations. Without batch normalization, the ReLU activations of the Li-GRU are unbounded and tend to cause numerical instabilities.  According to our experience, the convergence of Li-GRU without batch normalization, can be achieved only by setting rather small learning rate values. The latter setting, however, can lead to a poor performance and, as clearly emerged from this experiment, coupling  Li-GRU  with this technique is strongly recommended. 

\subsection{CTC results} \label{sec:cts}
Table \ref{tab:resctc} summarizes the results of CTC on the TIMIT data set. In these experiments, the Li-GRU clearly outperforms the standard GRU, showing the effectiveness of the proposed model even in a end-to-end ASR framework. The improvement is obtained both with and without batch normalization and, similarly to what observed for hybrid systems, the latter technique leads to better performance when coupled with Li-GRU. However, a smaller performance gain is observed when batch normalization is applied to the CTC. This result could also be related to the different choice of the regularizer, as weight noise was used instead of recurrent dropout. 

In general, PERs are higher than those of the hybrid systems. End-to-end methods, in fact, are relatively young models, that are currently still less competitive than more complex (and mature) DNN-HMM approaches. We believe that the gap between CTC and hybrid speech recognizers could be partially reduced in our experiments with a more careful setting of the hyperparameters and with the introduction of an external phone-based language model. The main focus of the paper, however, is to show the effectiveness of the proposed Li-GRU model, and a fair comparison between CTC and hybrid systems is out of the scope of this work. 

\begin{table}[t!]
\centering
\tabcolsep=0.25cm
    \begin{tabular}{ | l | c | c | }
    \cline{1-3}
   {\backslashbox{\em{Arch.}}{\em{Batch-norm.}}} & False &  True \\ \hline
GRU & 22.1 & 22.0 \\ \hline
Li-GRU & \textbf{21.1} & \textbf{20.9} \\ \hline
    \end{tabular}
\caption{PER(\%) obtained for the test set of TIMIT with various CTC RNN architectures.}
\label{tab:resctc}
\end{table}

\subsection{Other results on TIMIT} \label{sec:timit}
The results of  Table \ref{tab:bn} and \ref{tab:resctc} highlighted that the proposed Li-GRU model outperforms other GRU architectures. 
In this sub-section, we extend this study by performing a more detailed comparison with the most popular RNN architectures. To provide a fair comparison, batch normalization is hereinafter applied to all the considered RNN models. Moreover, at least five experiments varying the initialization seeds were conducted for each RNN architecture. The results are thus reported as the average PER with their corresponding standard deviation.

Table \ref{tab:res1} presents the ASR performance obtained with the TIMIT dataset. 
\begin{table}[t!]
\centering
\tabcolsep=0.20cm
    \begin{tabular}{ | l | c | c | c | c | }
    \cline{1-4}
   {\backslashbox{\em{Arch.}}{\em{Feat.}}} & MFCC &  FBANK & fMLLR \\ \hline
relu-RNN & 18.7 $\pm$ 0.18 & 18.3 $\pm$ 0.23 & 16.3  $\pm$ 0.11 \\ \hline
LSTM & 18.1 $\pm$ 0.33 & 17.1 $\pm$ 0.36 & 15.7  $\pm$ 0.32 \\ \hline
GRU & 17.1 $\pm$ 0.20 & 16.7 $\pm$ 0.36 & 15.3  $\pm$ 0.28 \\ \hline
M-GRU & 17.2 $\pm$ 0.11 & 16.7 $\pm$ 0.19 & 15.2  $\pm$ 0.10 \\ \hline
Li-GRU & \textbf{16.7} $\pm$ 0.26 & \textbf{15.8} $\pm$ 0.10 & \textbf{14.9}  $\pm$ 0.27
\\ \hline  
    \end{tabular}
\caption{PER(\%) obtained for the test set of TIMIT with various RNN architectures.}
\label{tab:res1}
\end{table}
The first row reports the results achieved with a simple RNN with ReLU activations (no gating mechanisms are used here). Although this architecture has recently shown promising results in some machine learning tasks \cite{orth_init}, our results confirm that gated recurrent networks (rows 2-5) outperform traditional RNNs.
We also observe that GRUs tend to slightly outperform the LSTM model.
As expected, M-GRU (i.e., the architecture without reset gate) achieves a performance very similar to that obtained with standard GRUs, further supporting our speculation on the redundant role played by the reset gate in a speech recognition application. 
The last row reports the performance achieved with the proposed model, in which, besides removing the reset gate, ReLU activations are used. 
The Li-GRU performance indicates that our architecture consistently outperforms the other RNNs over all the considered input features. A remarkable achievement is the average PER(\%) of $14.9$\% obtained with fMLLR features. To the best of our knowledge, this result yields the best published performance on the TIMIT test-set.

In Table ~\ref{tab:res_ph} the PER(\%) performance is split into five different phonetic categories (vowels, liquids, nasals, fricatives and stops), showing that Li-GRU exhibits the best results for all the considered classes.
 
\begin{table}[t!]
\centering
\tabcolsep=0.20cm
    \begin{tabular}{  | l | c | c | c | c | c |}
    \cline{1-4}
Phonetic Cat. & Phone Lists & GRU & Li-GRU  \\ \hline
Vowels & \{\textit{iy,ih,eh,ae,...,oy,aw,ow,er}\} & 23.2 & 23.0  \\ \hline
Liquids & \{\textit{l,r,y,w,el}\} & 20.1 & 19.0  \\ \hline
Nasals & \{\textit{en,m,n,ng}\} & 16.8 & 15.9 \\ \hline
Fricatives & \{\textit{ch,jh,dh,z,v,f,th,s,sh,hh,zh}\} & 17.6 & 17.0 \\ \hline
Stops & \{\textit{b,d,g,p,t,k,dx,cl,vcl,epi}\} & 18.5 & 17.9 \\ \hline  
    \end{tabular}
\caption{PER(\%) of the TIMIT dataset (MFCC features) split into five different phonetic categories. Silences (\textit{sil}) are not considered here.}
\label{tab:res_ph}
\end{table}

Previous results are obtained after optimizing the main hyperparameters of the model on the development set. Table \ref{tab:opt} reports the outcome of this optimization process, with the corresponding best architectures obtained for each RNN architecture.
 \begin{table}[t!]
 \centering
 \tabcolsep=0.25cm
     \begin{tabular}{  | l | c | c | c | c | c |}
     \cline{1-4}
 Architecture & Layers & Neurons & \# Params  \\ \hline
 relu-RNN & 4 & 607 & 6.1 M   \\ \hline
 LSTM & 5 & 375 & 8.8 M   \\ \hline
 GRU & 5 & 465  & 10.3 M  \\ \hline
 M-GRU & 5 & 465 & 7.4 M  \\ \hline
 Li-GRU & 5 & 465 & 7.4 M  \\ \hline  
     \end{tabular}
 \caption{Optimal number of layers and neurons for each TIMIT RNN model. The outcome of the optimization process is similar for all considered features.}
 \label{tab:opt}
 \end{table}
For GRU models, the best performance is achieved with 5 hidden layers of 465 neurons. It is also worth noting that M-GRU and Li-GRU have about 30\% fewer parameters compared to the standard GRU.

\subsection{Recognition performance on DIRHA English WSJ} \label{sec:dirha}
After a first set of experiments on TIMIT, in this sub-section we assess our model on a more challenging and realistic distant-talking task, using the  DIRHA English WSJ corpus. A challenging aspect of this dataset is the acoustic mismatch between training and testing conditions. Training, in fact, is performed with a reverberated version of WSJ, while test is characterized by both non-stationary noises and reverberation. 

Tables \ref{tab:res2} and \ref{tab:res3} summarize the results obtained with the simulated and real parts of this dataset. 
\begin{table}[t!]
\centering
\tabcolsep=0.30cm
    \begin{tabular}{ | l | c | c | c | c | }
    \cline{1-4}
   {\backslashbox{\em{Arch.}}{\em{Feat.}}} & MFCC &  FBANK & fMLLR \\ \hline
relu-RNN & 23.7  $\pm$ 0.21  & 23.5 $\pm$ 0.30 & 18.9 $\pm$ 0.26 \\ \hline
LSTM & 23.2  $\pm$ 0.46  & 23.2 $\pm$ 0.42  & 18.9 $\pm$ 0.24 \\ \hline
GRU & 22.3  $\pm$ 0.39   & 22.5 $\pm$ 0.38 & 18.6 $\pm$ 0.23 \\ \hline
M-GRU & 21.5  $\pm$ 0.43   & 22.0 $\pm$ 0.37  & 18.0 $\pm$ 0.21 \\ \hline
Li-GRU & \textbf{21.3}  $\pm$ 0.38  & \textbf{21.4} $\pm$ 0.32 & \textbf{17.6} $\pm$ 0.20 \\ \hline  
\end{tabular}
\caption{Word Error Rate (\%) obtained with the DIRHA English WSJ dataset (simulated part) for various RNN architectures.}
\label{tab:res2}
\end{table}
\begin{table}[t!]
\centering
\tabcolsep=0.30cm
    \begin{tabular}{ | l | c | c | c | c | }
    \cline{1-4}
   {\backslashbox{\em{Arch.}}{\em{Feat.}}} & MFCC &  FBANK & fMLLR \\ \hline
relu-RNN & 29.7 $\pm$ 0.31  & 30.0 $\pm$ 0.38 & 24.7 $\pm$ 0.28  \\ \hline
LSTM & 29.5 $\pm$ 0.41  & 29.1 $\pm$ 0.42 & 24.6 $\pm$ 0.35 \\ \hline
GRU & 28.5 $\pm$ 0.37  & 28.4 $\pm$ 0.21 & 24.0 $\pm$ 0.27 \\ \hline
M-GRU & 28.4 $\pm$ 0.34  & 28.1 $\pm$ 0.30 & 23.6 $\pm$ 0.21 \\ \hline
Li-GRU & \textbf{27.8} $\pm$ 0.38  & \textbf{27.6} $\pm$ 0.36 & \textbf{22.8} $\pm$ 0.26 \\ \hline  
\end{tabular}
\caption{Word Error Rate (\%) obtained with the DIRHA English WSJ dataset (real part) for various RNN architectures.}
\label{tab:res3}
\end{table}

\begin{figure}[t]
\centering
  \includegraphics[scale=0.50]{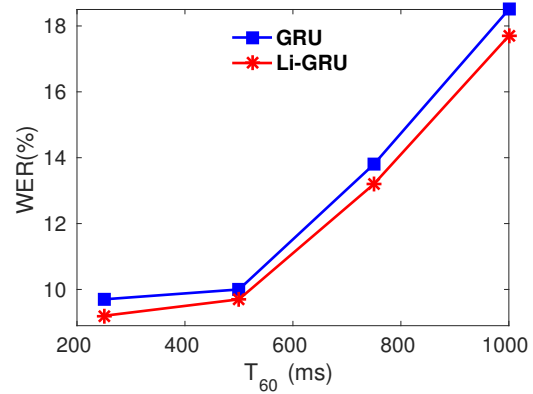}
\caption{Evolution of the WER(\%) for the DIRHA WSJ simulated data over different reverberation times $T_{60}$. }\label{fig:t60}
\end{figure}
These results exhibit a trend comparable to that observed for TIMIT, confirming that Li-GRU still outperform GRU even in a more challenging scenario. The results are consistent over both real and simulated data as well as across the different features considered in this study.

The reset gate removal seems to play a more crucial role in the addressed distant-talking scenario.  If the close-talking performance reported in Table \ref{tab:res1} highlights comparable error rates between standard GRU and M-GRU, in the distant-talking case we even observe a  small performance gain when removing the reset gate. We suppose that this behaviour is due to reverberation, that implicitly introduces redundancy in the signal, due to the multiple delayed replicas of each sample. This results in a forward memory effect, that can make reset gate ineffective. 

In Fig.~\ref{fig:t60}, we extend our previous experiments by generating simulated data with different reverberation times $T_{60}$ ranging from 250 to 1000 ms, as outlined in Sec. \ref{sec:corpora}. In order to simulate a more realistic situation, different impulse responses have been used for training and testing purposes. No additive noise is considered for these experiments.

As expected, the performance degrades as the reverberation time increases. Similarly to previous achievements, we still observe that Li-GRU outperform GRU under all the considered reverberation conditions. 

\subsection{Recognition performance on CHiME} \label{sec:chime}
In this sub-section we extend the results to the CHiME corpus, that is an important benchmark in the ASR field, thanks to the success of CHiME challenges \cite{chime,chime3}. In Tab.~\ref{tab:chime} a comparison across the various GRU architectures is presented. For the sake of comparison, the results obtained with the official CHiME 4 are also reported in the first two rows\footnote{The results obtained in this section are not directly comparable with the best systems of the CHiME 4 competition. Due to the purpose of this work, indeed, techniques such as multi-microphone processing, data-augmentation, system combination as well as lattice rescoring are not used here.}. 

\begin{table}[t!]
\centering
\tabcolsep=0.10cm
    \begin{tabular}{ | l | c | c | c | c | }
    \cline{1-5}
   {\backslashbox{\em{Arch.}}{\em{Dataset}}} & DT-sim & DT-real &  ET-sim & ET-real \\ \hline
DNN & 17.8 $\pm$ 0.38 & 16.1 $\pm$ 0.31  & 26.1 $\pm$ 0.45 & 30.0 $\pm$ 0.48 \\ \hline   
DNN+sMBR & 14.7 $\pm$ 0.25  & 15.7 $\pm$ 0.23  & 24.0 $\pm$ 0.31 & 27.0 $\pm$ 0.35 \\ \hline
GRU & 15.8 $\pm$ 0.30  & 14.8 $\pm$ 0.25  & 23.0 $\pm$ 0.38 & 23.3 $\pm$ 0.35 \\ \hline
M-GRU & 15.9 $\pm$ 0.32 & 14.1 $\pm$ 0.31  & 22.8 $\pm$ 0.39 & 23.0 $\pm$ 0.41 \\ \hline
Li-GRU & \textbf{13.5} $\pm$ 0.25  & \textbf{12.5} $\pm$ 0.22  & \textbf{20.3} $\pm$ 0.31 & \textbf{20.0} $\pm$ 0.33 \\ \hline
 
\end{tabular}
\caption{Speech recognition performance on the CHiME dataset (single channel, fMLLR features).}
\label{tab:chime}
\end{table}

Results confirm the trend previously observed, highlighting a significant relative improvement of about 14\%  achieved when passing from GRU to the proposed Li-GRU. Similarly to our findings of the previous section, some small benefits can be observed when removing the reset gate. The largest performance gap, however, is reached when adopting ReLU units (see M-GRU and Li-GRU columns), confirming the effectiveness of this architectural variation. Note also that the GRU systems significantly outperform the DNN baseline, even when the latter is based on sequence discriminative training (DNN+sMBR)\cite{sequence_training}.  


Tab.~\ref{tab:chime2} splits the ASR performance of the real test set into the four noisy categories.
\begin{table}[t!]
\centering
\tabcolsep=0.30cm
    \begin{tabular}{ | l | c | c | c | c | }
    \cline{1-5}
   {\backslashbox{\em{Arch.}}{\em{Env.}}} & BUS &  CAF & PED & STR \\ \hline
DNN & 44.1 & 32.0 & 26.2 & 17.7 \\ \hline
DNN+sMBR & 40.5 & 28.3 & 22.9 & 16.3 \\ \hline
GRU & 33.5 & 25.6 & 19.5 & 14.6 \\ \hline
M-GRU & 33.1 & 24.9 & 19.2 & 14.9 \\ \hline
Li-GRU & \textbf{28.0} & \textbf{22.1} & \textbf{16.9} & \textbf{13.2} \\ \hline
 
\end{tabular}
\caption{Comparison between GRU and Li-GRU for the four different noisy conditions considered in CHiME on the real evaluation set (ET-real).}
\label{tab:chime2}
\end{table}
Li-GRU outperforms GRU in all the considered environments, with a  performance gain that is higher when more challenging acoustic conditions are met. For instance, we obtain a relative improvement of 16\% in the bus (BUS) environment (the noisiest), against the relative improvement of 9.5\% observed in the street (STR) recordings.

\subsection{Recognition performance on TED-talks} 
Tab. \ref{tab:ted_talks} reports a comparison between GRU and Li-GRU on the TED-talks corpus. The experiments are performed with standard MFCC features, and a four-gram language model is considered in the decoding step (see \cite{ted_lm} for more details).

Results on both test sets consistently shows the performance gain achieved with the proposed architecture. This further confirms the effectiveness of Li-GRU, even for a larger scale ASR task. In particular, a relative improvement of about 14-17\% is achieved. This improvement is statistically significant according to the Matched Pairs Sentence Segment Word Error Test (MPSSW) \cite{statistical_significance}, that is conducted with NIST sclite $sc\_sta$t tool with a p-value of $0.01$.

\label{sec:ted_talks}
\begin{table}[t!]
\centering
\tabcolsep=0.30cm
    \begin{tabular}{ | l | c | c | c | c | }
    \cline{1-3}
   {\backslashbox{\em{Arch.}}{\em{Dataset.}}} & TST-2011 &  TST-2012  \\ \hline
GRU & 16.3 $\pm$ 0.13 & 17.0 $\pm$ 0.16 \\ \hline
Li-GRU & \textbf{13.8} $\pm$ 0.12 & \textbf{14.4} $\pm$ 0.15 \\ \hline
 
\end{tabular}
\caption{Comparison between GRU and Li-GRU with the TED-talks corpus.}
\label{tab:ted_talks}
\end{table}

\subsection{Training time comparison} \label{sec:tr_time}
In the previous subsections, we reported several speech recognition results, showing that Li-GRU outperforms other RNNs. In this sub-section, we finally focus on another key aspect of the proposed architecture, namely its improved computational efficiency. 
In Table \ref{tab:tr_time}, we compare the per-epoch wall-clock training time of GRU and Li-GRU models. 

\begin{table}[t!]
\centering
\tabcolsep=0.20cm
    \begin{tabular}{ | l | c | c | c | c | }
    \cline{1-5}
   {\backslashbox{\em{Arch.}}{\em{Dataset.}}} & TIMIT &  DIRHA & CHiME & TED \\ \hline
GRU & 9.6 min & 40 min & 312 min & 590 min\\ \hline
Li-GRU & \textbf{6.5 min}   & \textbf{25 min} & \textbf{205 min} & \textbf{447 min} \\ \hline  
\end{tabular}
\caption{Per-epoch training time (in minutes) of GRU and Li-GRU models for the various datasets on an NVIDIA K40 GPU.}
\label{tab:tr_time}
\end{table}

The training time reduction achieved with the proposed architecture is about 30\% for all the datasets. This reduction reflects the amount of parameters saved by Li-GRU, that is also around 30\%. The reduction of the computational complexity, originated by a more compact model, also arises for testing purposes, making our model potentially suitable for small-footprint ASR, \cite{small1,small2,small3,small4,small5,online2}, which studies DNNs designed for portable devices with small computational capabilities. 

\section{Conclusions} \label{sec:conc}
In this paper, we revised standard GRUs for speech recognition purposes. The proposed Li-GRU architecture is a simplified version of a standard GRU, in which the reset gate is removed and ReLU activations are considered. Batch normalization is also used to  further improve the system performance as well as to limit the numerical instabilities originated from ReLU non-linearities.

The experiments, conducted on different ASR paradigms, tasks, features and environmental conditions, have confirmed the effectiveness of the proposed model. The Li-GRU, in fact, not only yields a better recognition performance, but also reduces the computational complexity, with a reduction of more than 30\% of the training time over a standard GRU.

Future efforts will be focused on extending this work to other speech-based tasks, such as speech enhancement and speech separation as well as to explore the use of Li-GRU in other possible fields.


 \section*{Acknowledgments}
 The authors would like to thank Piergiorgio Svaizer for his insightful suggestions on an earlier version of this paper. We would also thank the anonymous reviewers for their careful reading of our manuscript and their helpful comments.

We gratefully acknowledge the support of NVIDIA Corporation with the donation of a Tesla K40 GPU used for this research. Computations were also made on the Helios supercomputer from the University of Montreal, managed by Calcul Québec and Compute Canada.

\bibliographystyle{IEEEtran}
\bibliography{mybibfile}


\end{document}